\documentstyle{article}

\begin{document}

\begin{center}
{\huge\bf A Two dimensional Model of Superconductivity}
\end{center}

\vspace{1cm}
\begin{center}
{\large\bf 
F. GHABOUSSI}\\
\end{center}

\begin{center}
\begin{minipage}{8cm}
Department of Physics, University of Konstanz\\
P.O. Box 5560, D 78434 Konstanz, Germany\\
E-mail: ghabousi@kaluza.physik.uni-konstanz.de
\end{minipage}
\end{center}

\vspace{1cm}

\begin{center}
{\large{\bf Abstract}}
\end{center}

\begin{center}
\begin{minipage}{12cm}
We present a two dimensional model of superconductivity where bosonization of fermions is described by topological fermion-boson duality. The model solves the discrepancy between theoretical and empirical values of penetration depth and explains the appearence of quantized vortex lines in supercunductors in accord with cyclotron motion of supercunducting electrons.
\end{minipage}
\end{center}

\newpage
First let us recall that superconductivity is defined by two basic properties of superconductors: The absence of usual resistivity relation in view of a new relation between electromagnetic field and current which is represented by London equations; and the appearence of quantized magnetic flux in supercunductor, that expels the applied magnetic field. We show that both properties can be explained by a {\it two dimensional} microscopic (quantum) 
model of supercunductivity, where the phenomenological London equations \cite{pheno} are derived as equations of motion; and the flux quantization results from the {\it canonical quantization} of theory. Thus flux quantization is an invariant {\it two dimensional} concept, in view of its definition with respect to {\it surface} and in view of appearence of flux quantization on the {\it two dimensional} quantum Hall samples relating supercunductivity with  quantum Hall effect \cite{prang}, which we presented recently a purely {\it two dimensional} model for \cite{me}. In this sense, also the supercunductivity can be considered as a two dimensional effect described by a two dimensional model. Thus also other theoretical models of supercunductivity, i. e. Ginzburg-Landau model, which is conceived four dimensionally, is based on two dimensional fundations (see below). Thus following investigations of theoretical and empirical aspects of superconductivity show that basic phenomenological concepts of supercunductivity, are in fact two dimensional concepts which can be given by a two dimensional theory:

With respect to theoretical aspects, note that the solutions of combined Maxwell-London equations: $\Delta B_i = \displaystyle{\frac{1}{\lambda^2}} B_i$, are given by: $B_i = B_{i (surface)} e^{- \frac{X}{\lambda}}$, where $\lambda$ is London penetration depth and $B_{i (surface)}$ is the magnetic field strength on the {\it two dimensional} surface of supercunductor. Hence the solution $B_i$, is dominated by the {\it surface} solution. This shows the {\it two dimensional} structure of supercunductivity, even if it is applied three and four dimensional London- and Maxwell equations. Further note that the Ginzburg-Landau equations are accompanied by a boundary condition: $(- i \hbar \partial_3 - e A_3 ) \psi = 0$ which allows electronic currents {\it on the surface only}. Therefore also in Ginzburg-Landau model, the electronic dynamics which causes the supercundutivity, is restricted to the electronic dynamics on the surface of supercundutors only.

Note that also in the standard approach to supercunductivity, the so called Fermi {\it surface} where the electrons motion takes place, is in any case a {\it surface} which can be described by two independent variables only. Thus the motion of electrons on such a momentum surface is a {\it two dimensional} motion \cite{buch}. We show that such a two dimensional momentum space is, just in view of its two dimensionality, equivalent to a two dimensional configuration space. Further note that the exact solution of corresponding {\it two dimensional} Schroedinger equation results in energy levels with energy gap \cite{buch}. In other words, also the phenomenological concept of energy gap, is, from point of view of Schroedinger theory, equivalent to a two dimensional concept (see also below). These facts show the {\it two dimensional} basis of standard approach to supercunductivity, although it is conceived as a three- or four dimensional approach.

We show also that our two dimensional theory is equivalent by Stokes theorem to a one dimensional theory on the contour region of two dimensional superconducting sample, hence the bosonization of fermions can be considered in this model, as a result of Hodge duality between fermions and bosons on the one dimensional contour manifold \cite{nak}. We show further that flux quantization with respect to electrons, is equivalent to the quantization of cyclotron motion of electrons; hence in this model the quantized vortex lines are due to quantized cyclotronic currents of electrons (see below). 

Furthermore for dimensional structure of supercunductivity in view of flux quantization, note that the flux quantization integral: $e \int \int\limits_{surface} F_{lm} d X^l \wedge dX^m = N h \ , \ N \in {\mathbf Z}$,  is a {\it two dimensional} topologically invariant quantity: Since its invariant value $(N h)$, is defined with respect to a two dimensional {\it surface} with surface element $(d X^l \wedge dX^m)$. Therefore the invaraint defintion of flux is entirely a two dimensional concept and the flux quantization is a two dimensional effect. Then the supercunductivity which is defined by flux quantization, can be understood also as a two dimensional effect. 

With respect to empirical results note that, it is known that there is a remarkable discrepancy between the measured values and the theoretical value of London penetration depth given by: $\lambda^2 := \displaystyle{\frac{M_s}{\mu_0 n_s e_s ^2}}$ where $M_s$, $e_s$ and $n_s$ are mass, charge and {\it volume density} of Cooper pairs. This fact shows that something is wrong with the structure of London model, although it is a good phenomenological model \cite{pheno}. Thus it is also a strong hint to choose a new structure for London equations, where the peneteration depth can be defined in a manner which fits to its empirical values. We show, that the mentioned discrepancy can be corrected, if one uses the conjectured {\it two dimensional} model where the penetration depth is defined with respect to a {\it two dimensional surface} density.

Note also that, the "two dimensionality" of supercunducting {\it rings}, is an empirical hint for the main role played by two dimensional structures in supercunductivity. Thus the two dimensional tin films play very importent role in application of supercunductivity, specially in supercunducting electronics \cite{buch}. 

Taking all these theoretical and empirical facts about the main role played by the two dimensional system of electrons in supercunductivity into account; one is halten to consider the supercunductivity as caused by the two dimensional motion of electrons in a magnetic field, under certain conditions. In other words we argue that, in view of mentioned empirical and theoretical difficulties of three dimensional model of supercunductivity, and in view of several accordencies of empirical and phenomenological aspects of supercunductivity with {\it two dimensional} concepts: The supercunductivity can be considered as an effect which results from the quantum behaviour of electrons on the {\it surface} of supercunductors, interacting with  electromagnetic field. Thus, we show that supercunducting effects which are described by London equations and flux quantization, follow from the quantum electrodynamical behaviour of electrons with {\it two degrees of freedom}. Thus the theory of supercunductivity can be formulated, as the {\it two dimensional} quantum electrodynamics of electrons with two degrees of freedom only. 

The action of supercunductivity is given by the two dimensional topologically invariant action of an electron interacting with electromagnetic potential $A_m$ or with magnetic field strengh $F_{ml}$, in the single electron picture \cite{non}:

\begin{equation}
S = \displaystyle{\frac{1}{2}} (\int \int\limits_{surface} dP_m \wedge d X^m + e \int \int\limits_{surface} F_{lm} d X^l \wedge dX^m) = \displaystyle{\frac{1}{2}} (\oint\limits_{contour} P_m d X^m + e \oint\limits_{contour} A_m d X^m) \ \ ,
\end{equation} 

where $P_m$ and $X_m; \ , l, m = 1, 2$ are the momentum and the position coordinates of an electron and the {\it surface} integral is considered over the surface of supercunductor, whereas the equivalent {\it contour} integral is considered on the contour region of superconductor surface. Thus the equality represents the Stokes theorem for both kinetic and petential term of electron.

Here note that the {\it actual motion} of a physical system takes place always on a polarized phase space which contains the half of phase space variables \cite{geq}, thus the action function of a system and its wave funtion are always functions of half of phase space variable, beside the time parameter. The best example of such a polarized phase space with half of phase space variables is the configuration space which is known, in the case that the wave function is a function on this space, as the position representation of wave function. The configuration space of supercunducting system is the two dimensional surface where the position of electrons are defined. Therefore the {\it actual} action function of the two dimensional supercunductivity can be defined on the polarized phase space of the supercunducting system, i. e. on the two dimensional configuration space of the system, or on the two dimensional surface of supercunductor. It is in this manner that we can write the action function (1) on the surface of supercunductor, since the surface represents the polarized phase space of our supercunducting system. 

Nevertheless the contour action (1) can be rewritten by the current density $J_m := n_e e \displaystyle{\frac{P_m}{M_e}}$ as:

\begin{equation}
S = \displaystyle{\frac{1}{2}} (\displaystyle{\frac{M_e}{n_e e}} \oint\limits_{contour} J_m d X^m + e \oint\limits_{contour} A_m d X^m) \ \ ,
\end{equation} 

where $n_e$ is the homogenous or spatially constant volume density of electrons. Note that despite of assertion that Ginzburg-Landau theory considers a variable density $n (r) := | \psi (r) |^2$, actually 

$n (r) := | \psi (r) |^2$ does not play the role of a true variable in this theory, since the true variables here are $\psi (r)$, $A_i$ only, with respect to which the action of theory is varied in order to obtain the equations of motion of theory: Whereas the $n (r) := | \psi (r) |^2$ is not varied in this theory at all. Thus in contrary, the actual value of density in this theory is given by: $| \psi (r) |^2 = - \displaystyle{\frac{\alpha}{\beta}}$ in the thermodynamical equilibrium, where $\alpha$ and $\beta$ are dimensional constants, since also they are not varied in the theory. Otherwise the theory would possess further equations of motion with respect to variation of $\alpha$, $\beta$ or $| \psi (r) |^2$. Whereas the whole dynamics of Ginzburg-Landau model is described by the two equations of motion which result from the variation of action with respect to $\psi (r)$, $A_i$ variables. In other words, although $\psi (r)$ is a variable, nevertheless $| \psi (r) |^2$ is constant in this theory, in accord with the constancy of $\alpha$ and $\beta$ in $| \psi (r) |^2 = - \displaystyle{\frac{\alpha}{\beta}}$.
Recall that also the coherence length is obtained in this theory in accord with $| \psi (r) |^2 = - \displaystyle{\frac{\alpha}{\beta}}$. Therefore even in the Ginzburg-Landau model of supercunductivity the density of electrons or Cooper pairs is actually {\it constant}. 

Note further that in view of the equivality: $\displaystyle{\frac{M_e}{n_e e^2}} = \displaystyle{\frac{M_s}{n_s e_s ^2}}$ where the index s denotes Cooper pairs, the action (2) is also valid for Cooper pair of electrons. Thus the two dimensional model applies also as a Cooper pair model, if one replaces $J_m$, $X_m$, and $\displaystyle{\frac{M_e}{n_e e}}$ by $J_m ^s$, $X_m ^s$ and $\displaystyle{\frac{M_s}{n_s e_s}}$, respectively.

\bigskip
On the {\it surface} of supercunductor, the two position variables of electron, i. e. $X^l$, are the actual variables of supercunducting system of electrons. Therefore the Euler-Lagrange equations of system (2): $\displaystyle{\frac{\partial L}{\partial X^l}} = \partial_m \displaystyle{\frac{\partial L}{\partial \partial_m X^l}} = 0$ are given for a variation of $S$ with respect to $X^l$ by:

\begin{equation}
\epsilon_{lm} \partial_l J_m = - \displaystyle{\frac{1}{\mu_0 \lambda^2}} B \ \ \ \epsilon_{lm}:= - \epsilon_{ml} = 1 \ \ \,
\end{equation}

, in accord with $d X^l := \displaystyle{\frac{\partial X^l}{\partial X^m}} d X^m = \partial_m X^l d X^m$ and $\displaystyle{\frac{\partial L}{\partial X^l}} = 0$, where $B := \epsilon_{lm} \partial_l A_m$ and $\lambda$ is the London penetration depth: $\lambda^2 := \displaystyle{\frac{M_e}{\mu_0 n_e e^2}}$ which is defined here for electrons. Note that $\displaystyle{\frac{\partial X^l}{\partial X^m}}$ is in general not constant, i. e. for a cyclotronic motion with $X^1 := r \ cos \alpha$ and $X^2 := r \ sin \alpha$, or also on a curved manifold, it is not constant.

Nevertheless, in accord with $(dX^l = \dot{X}^l dt)$, one may consider the time parameter $t$, instead of $X^l$, as the variable of system. Then the equations of motion of system are given by:

\begin{equation}
\partial_t {J}_m = \displaystyle{\frac{1}{\mu_0 \lambda^2}} E_m  \ \ \,
\end{equation}

where $E_m := - \partial_t A_m$.

These are London equations which are derived here as equations of motion from a two dimensional action function (1) = (2). It shows that the theory of supercunductivity in accord with London equations, can be described by a two dimensional model for electrons with only two degrees of freedom.

We show now the flux quantization, as a result of canonical quantization of action (1) in the sense of Bohr-Sommerfeld quantization \cite{canon}:

The action (1) can be rewritten also by:

\begin{equation}
S = \int \int\limits_{surface} d \pi_m \wedge d X^m  = \oint\limits_{contour} \pi_m d X^m \ \ ,
\end{equation}

where $\pi_m := \displaystyle{\frac{1}{2}} ( P_m + e A_m )$.

Then, Bohr-Sommerfeld quantization of this system is given by:

\begin{equation}
S = \int \int\limits_{surface} d \pi_m \wedge d X^m = \oint\limits_{contour} \pi_m d X^m  = N h \ \ , N \in {\mathbf Z},
\end{equation}

or by: 

\begin{equation}
S_1 = \int \int\limits_{surface} d P_m \wedge d X^m = \oint\limits_{contour} P_m d X^m = N h \ \ ,
\end{equation}

and

\begin{equation}
S_2 = \int \int\limits_{surface} F_{lm} d X^l \wedge dX^m) =  \oint\limits_{contour}  A_m d X^m) = N \displaystyle{\frac{h}{e}}
\end{equation}

The Bohr-Sommerfeld quantization postulate (7) is equivalent to the commutator quantization: 

$[ \hat{P}_m , \hat{X}_m] = - i \hbar$ \cite{canon}, describing the quantum behaviour of electron, whereas the Bohr-Sommerfeld quantization (8) describes the flux quantization in quantum units of $( \displaystyle{\frac{h}{e}} )$. Thus flux quantization can be considered as the canonical quantization of present two dimensional electrodynamics. Thus Bohr-Sommerfeld quantization (8) is equivalent to the commutator quantization $e [ \hat{A}_m , \hat{X}_m] = - i \hbar$ \cite{canon}, describing the quantum behaviour of electromagnetic potential with respect to electrons (see also below). By a comparision of these commutators: $[ \hat{P}_m , \hat{X}_m] = e [ \hat{A}_m , \hat{X}_m] = - i \hbar$, one obtains: $\hat{P}_m = e \hat{A}_m$, which recalls either the flux quantization condition of vanishing of velocity of electrons: $V_m = (M_e )^{-1} (P_m - e A_m ) = 0$ on the region of contour integration in the standard approach, or the boundary condition of Ginzburg-Landau theory.

For bosonization note that, the topological invariance of flux quantization is enough hint about the importent role played by topology in supercunductivity. Thus, as we show, the bosonization of electrons is a topological property of two dimensional model with boundary: Since in this case, as it is obvious from the action (1), the two dimensional model is equivalent to a one dimensional model on the one dimensional contour region. In order to investigate the standard topology of this model, one has to use the differential form representation of fermions and bosons. Then we attach odd differential forms to fermions and even differential forms to bosons, as the usual attachment \cite{wit}. Therefore, in our one dimensional model which is represented by the contour action on the one dimensional contour manifold, the fermions are considered as one-forms $\Omega^1 := \Omega_m d X^m$ which obey the Fermi statistics in accord with exterior algebra of forms: $[ \Omega^1 _1 \ , \ \Omega^1 _2 ]_+ = 0$, in view of exterior algebra: $d X^l \wedge d X^m + d X^m \wedge d X^l = 0$. Further we consider bosons in our one dimensional case, as even, i. e. zero-forms $\Omega^0$ or scalar functions which obey the Bose statistics in accord with: $[ \Omega^0 _1 \ , \ \Omega^0 _2 ]_- = 0$, since zero forms commute always. In other words we have the following mathematically well defined differential form representions for fermions $(f_m \in \Omega^1)$ and bosons $(b_m \in \Omega^0)$ in our one dimensional case, which obey: $[ \Omega^1 _1 \ , \ \Omega^1 _2 ]_+ = [ f_1 \ , \ f_2 ]_+ = 0$ or $f_1 f_2 = - f_2 f_1$ and $[ \Omega^0 _1 \ , \ \Omega^0 _2 ]_- = [ b_1 \ , \  b_2 ]_- = 0$ or $b_1 b_2 = b_2 b_1$, in accord with the standard statistics.

On the other hand, as on any manifold, also on the one dimensional contour manifold where electrons are concentrated in this model, there exists the so called Hodge duality between differential forms. In other words, on a m-dimensional manifold a r-form $\Omega^r$ is dual to a $(m - r)$-form $\Omega^{m - r}$, i. e.: $\Omega^r = * \Omega^{m - r}$ \cite{nak}. 
Hence on a one-dimensional manifold one-forms are dual to zero-forms: $\Omega^0 = * \Omega^1$: In other words on the one dimensional contour manifold of supercunductor, zero-form bosons and one-form fermions are dual to each other, i. e. $b_m = * f_m$. Therefore fermions obey on the supercunducting contour manifold $(1D)$ the Bose statistics, in view of their Hodge duality with bosons; since by $b_m = * f_m$ the Bose statistics relation: $[ b_1 \ , \  b_2 ]_- = 0$ can be rewritten by: $[ f_1 \ , \  f_2 ]_- (1 D) = 0$, in view of commutativity of boson zero forms with fermion one forms and $*^2 \equiv 1$: Thus, in view of the absence of two forms on a one dimensional manifold and the fact that the commutator of one forms are two forms, i. e. in view of $[ \Omega^1 \ , \ \Omega^1 ]_- \in \Omega^2$, the commutator of fermion one forms should vanish on the one dimensional contour manifold: $(1 D)$, i. e. $[ \Omega^1 \ , \ \Omega^1 ]_- ( 1 D) \equiv 0$; so that they should obey the Bose statistics on the contour manifold: $[ f_1 \ , \ f_2 ]_- (1 D) = 0$ or $( f_1 f_2 = f_2 f_1 )_{(1 D)}$. This is the prove of Bose statistics of fermions in the contour manifold.
Therefore electrons as fermion one forms on the contour manifold, obey the Bose statistics and behave themselves as bosons, occupaying the ground state collectively in very low temperatures. This explains the reason for the bosonization of electrons on the contour region of supercunductors. 

Note further that the quantum state of electrons in supercunductivity is manifested by flux quantization which is accompanied, as we showed with quantization of total energy of electron, as the sum of quantized kinetic energy: $\oint\limits_{contour} P_m d X^m = \oint\limits_{contour} P_m \dot{X}^m d t = N h$ and the quantized potential energy: $e \oint\limits_{contour} A_m d X^m = N h$. Thus we have by quantization of action (1) the quantization of energy levels of electron. This means that the ground state and the excited states of electron are quantized and separated from each other, in accord with the separation of {\it allowed energy levels}. In other words, even if the kinetic energy vanishes at zero temperature, the existence of {\it potential} and the quantization of {\it potential energy} in this model, quantizes the electron energy and separates the ground state of electron from excited states. Thus, the enegy spectrum of electron in certain {\it potentials} possesses energy gaps \cite{buc2}. Hence, in this model, there exists always, i. e. even at zero temperature, an {\it energy gap} between the ground state and the excited states, in accord with the quantized {\it potential energy}. 

Note also that, the formula for energy gap is the same as the formula for the energy levels of bound state of a quantized {\it two dimensional} system in Schroedinger theory \cite{buch}. Thus also in the standard approach, the electrons on the Fermi {\it surface} are executing essentially {\it two dimensional motion} (in momentum space) \cite{buch}. Nevertheless, just in the two dimensional case, the momentum is given by: $P_m = e A_m = e B \cdot X^l \epsilon_{ml}$, in accord with above arguments and in view of constancy of $B$ in this case: Thus, $A_m = e B \cdot X^l \epsilon_{ml}$ is the general definition of two dimensional electromagnetic potential on a {\it two dimensional} manifold, up to a constant, in view of constancy of magnetic field strength on the two dimensional manifold \cite{twform}. Therefore, in the two dimensional case under consideration, the momentum space {\it variables} $P_m$ are replaced by the configuration space {\it variables} $X^l$, in view of {\it constancy} of $B$ in $P_m = e B \cdot X^l \epsilon_{ml}$; and the {\it two dimensional} Fermi surface of such a system becomes equivalent to the two dimensional configuration space of system. Therefore the two dimensional motion of electrons on the Fermi surface is indeed nothing else than the motion on the two dimensional configuration space of system, or on its {\it surface}. 

These facts underline the two dimensional nature of supercunductivity and the two dimensionality of {\it energy gap} conception which is used sometimes instead of concept of {\it separated energy levels} in quantized systems. 

Further the {\it two dimensional} model of supercunductivity can describe the appearence of quantizaed vortex lines in a supercunductor, in view of canonical equivalence of flux quantization and the cyclotron motion of electrons in the {\it two dimensional} model: 

Note that on the one hand, we may rewrite the quantized electromagnetic action $S_2$ in (8) by:

\begin{equation}
S_2 = B \int \int\limits_{surface} \epsilon_{lm} d X^l \wedge dX^m) =  \oint\limits_{contour}  A_m d X^m) = N \displaystyle{\frac{h}{e}} \ \ , 
\end{equation}

in view of constancy of magnetic field $B := \epsilon_{lm} F_{lm} = (constant)$ on the two dimensional manifold \cite{twform}. On the other hand, this quantization can be compared, as a canonical quantization, with the general formula for canonical quantization \cite{canon}:

\begin{equation}
S = \int \int d \Pi_l \wedge d Q_l = \oint \Pi_l d Q_l = N h, \ N \in {\mathbf Z} \ \ ,
\end{equation}

which is equivalent to the canonical quantum commutator: $[ \hat{\Pi}_l ,  \hat{Q}_m ] = -i \hbar \delta_{lm}$ \cite{canon}. By this comparison, i. e. by $\Pi_l = A_l = e B \cdot X^m \epsilon_{lm}$ and $Q_l = X_l$, it is obvious that the canonical quantization postulate (9) is equivalent to the commutator quantization postulate:

\begin{equation}
B [ \hat{X}_l ,  \hat{X}_m ] = -i \epsilon_{lm} \displaystyle{\frac{\hbar}{e}}  \ \ ,
\end{equation}

which is the defining commutator for the {\it two dimensional} cyclotron motion of electron in the magnetic field $B$. Thus the quantization units in flux quantization and cyclotron motion are also the same: $( \displaystyle{\frac{\hbar}{e}} )$. In this sense, the flux qunatization and the cyclotron motion are equivalent quantization relations for one and the same {\it two dimensional motion} of electrons in a constant magnetic field: Where the flux quantization is the topological or global description of such a magnetic quantization in accord with Bohr-Sommerfeld postulate for the electromagnetic field: $S_2 = B \int \int\limits_{surface} \epsilon_{lm} d X^l \wedge dX^m) =  \oint\limits_{contour}  A_m d X^m) = N \displaystyle{\frac{\hbar}{e}}$, whereas the cyclotron motion is the local description of the same magnetic quantization in accord with: $B [ \hat{X}_m ,  \hat{X}_l ] = -i \epsilon_{lm} \displaystyle{\frac{\hbar}{e}}$. In other words, as like as any other effect, the magnetic quantization has its local and global aspects which are manifested by cyclotron motion and flux quantization, respectively. Hence, in supercunductivity, the flux quantization is accompanied by the local cyclotron motion of electrons. Thus supercunducting electrons in a magnetic flux, execute a cyclotron motion which causes an opposite magnetic flux that compensates the other applied flux. 

Note that such a canonical equivalence between flux quantiaztion in supercunductivity and the cylotron motion of electrons, explains the appearence of vortex lines and their quantization in supercunductors in a canonical manner: If one considers the cyclotron currents of electrons in supercunductors, as vortex lines which are quantized in units of 
$( \displaystyle{\frac{\hbar}{e}} )$, in view of cyclotron commutator quantization: $B [ \hat{X}_m ,  \hat{X}_l ] = -i \epsilon_{lm} \displaystyle{\frac{\hbar}{e}}$, or in accord with the equivalent flux quantization. In other words the quantized vortex lines are due to quantized cyclotron motion of electrons or to flux quantization. Thus such an accordence of flux quantization with the two dimensional cyclotron motion manifests again the two dimensionality of flux quantization and supercunductivity.

In conclusion note that the {\it two dimensional model} of supercunductivity has the advantage to present a theoretical value for penetration depth which fits better to the empirical values of penetration depth, than the theoretical values of other models. In other words our model presents a reliable method to correct the discrepancy between empirical and theoretical values of penetration depth which appears in other models \cite{mod}: Thus considering the definition of penetration depth by: $\lambda^2 : = \displaystyle{\frac{M_s}{\mu_0 n_s e_s ^2}}$ in London and related models, it is obvious that in view of the fixed value of $\displaystyle{\frac{M_s}{\mu_0 e_s ^2}}$, the only magnitude which can be responsible for such a discrepancy is the value of density: $n_s$. Hence by the change of value of the density only, one can obtain a theoretical value of penetration depth that fits to its empirical values. Thus a change in the density value, can be reached by the circumstance that, in a two dimensional model of supercunductiuvity, only surface electrons and hence surface density of electrons is relevant. Note that the above definition of $\lambda$ in London and related models results just, in view of use of {\it three dimensional} volume density in these models, since $dim ( n_{s (London)}) = dim ( n_{(s 3)} ) = L^{-3}$, $dim ( M_s) = L^{-1}$, $dim (\mu_0)  = dim ( e_s) = L^0$ and $dim ( \lambda ) = L^1$, in geometric units; so that the quantity $( \displaystyle{\frac{M_s}{\mu_0 n_{(s 3)} e_s ^2}} )$ is here of dimension: $L^2$. Whereas in {\it two dimensional} model, the dimension of density of electrons is: $dim (n_{(s 2)} ) = dim (n_{(e 2)} ) = L^{- 2}$ and the quantity $( \displaystyle{\frac{M_s}{\mu_0 n_{(s 2)} e_s ^2}} )$ is of dimension: $L$. Thus, in a two dimensional model of supercunductivity, the penetration depth can be defined by $\lambda := \displaystyle{\frac{M_s}{\mu_0 n_{(s 2)} e_s ^2}}$ only, which is equal to $\lambda := \displaystyle{\frac{M_e}{\mu_0 n_{(e 2)} e^2}}$. This change fits the theoretical value of $\lambda$ to its empirical values: 
Recall that the empirical values for penetration depth are several times larger than the theoretical value given by $\lambda_{(3D)} : = ( \displaystyle{\frac{M_s}{\mu_0 n_s e_s ^2}} )^{\frac{1}{2}}$ in London and related models. Then, if one {\it estimates} the surface density by: $n_{(s 2)} = n_{(s 3)} ^{\frac{2}{3}}$ with respect to the volume density which is known to be about $10^{21}/ cm^3$ in experiments \cite{exp}. One obtains, in accord with the known value of $\displaystyle{\frac{M_s}{\mu_0 e_s ^2}}$, a theoretical value of penetration depth $\lambda := \displaystyle{\frac{M_s}{\mu_0 n_{(s 2)} e_s ^2}}$ in the two dimensional model, which is about ten times larger than its value in the three dimensional London model. Then the value of penetration depth in the two dimensional model fits to the empirical values of penetration depth which are {\it about} five times larger than those in the London model.


\begin{thebibliography}{100}
\bibitem{canon}
The canonical quantization of a system represented by the action $S = \int \int d \Pi_i \wedge d Q_i$, can be given in two equivalent ways: Either by posulating the quantization of action, in the Bohr-Sommerfeld sense, i. e. by: $S = \int \int d \Pi_i \wedge d Q_i = \oint \Pi_i d Q_i = N h, \ N \in {\mathbf Z}$. Or by postulating the quantum commutator for operators of canonically conjugate variables: $[ \hat{\Pi}_i ,  \hat{Q}_j ] = -i \hbar \delta_{ij}$.
\end{thebibliography}
\end{document}